\documentclass[%
reprint,
showpacs,preprintnumbers,
amsmath,amssymb,
aps,
]{revtex4-1}

\usepackage{tikz}
\usepackage{pgffor}
\usepackage{verbatim}
\usepackage{pstricks}
\usepackage{pst-3dplot}
\usepackage[colorlinks, breaklinks=true, linkcolor=purple, citecolor=purple, linktocpage=true]{hyperref}
\usepackage{color}
\usepackage{graphicx}
\usepackage{dcolumn}
\usepackage{bm}

\usepackage{txfonts}

\begin{document}

\title{Chiral gravitomagnetic effect in topological superconductors and superfluids}

\author{Akihiko Sekine}
%\email{sekine@imr.tohoku.ac.jp}
\affiliation{Institute for Materials Research, Tohoku University, Sendai 980-8577, Japan}

\date{\today}

\begin{abstract}
We theoretically search for dynamical cross-correlated responses of three-dimensional topological superconductors and superfluids.
It has been suggested that a gravitational topological term, which is analogous to the theta term in topological insulators, can be derived in three-dimensional time-reversal invariant topological superconductors and superfluids, and that the dynamical gravitational axion field can be realized by the fluctuation of the relative phase, i.e., by the Leggett mode between topological $p$-wave pairing and conventional $s$-wave pairing.
In the presence of the dynamical gravitational axion field, we propose the emergence of the ``chiral gravitomagnetic effect'', a thermal current generation by gravitomagnetic fields, i.e., by mechanical rotations.
This effect can be regarded as a thermal counterpart of the chiral magnetic effect which has been studied mainly in Weyl semimetals.
We also show the occurrence of the anomalous thermal Hall effect in the bulk.
We discuss a possible application of our study to the thermal responses of Weyl superconductors.
\end{abstract}


\maketitle

\section{Introduction}
Search for novel phases and phenomena is one of the fundamental research themes in condensed matter physics.
Recently, topological phases of matter such as topological insulators and superconductors have attracted great attention as novel phases \cite{Hasan2010,Qi2011,Ando2013}.
The low-energy effective models for these topological phases can be described by relativistic fermions, i.e., Dirac or Majorana fermions \cite{Hasan2010,Qi2011,Ando2013}.
Novel phenomena emerging from the existence of relativistic fermions have been intensively and extensively searched for.
In the presence of bulk energy gaps, topological phases are classified and distinguished from topologically trivial phases by the topological invariants \cite{Kane2005,Fu2007,Moore2007,Roy2009,Schnyder2008,Ryu2010}.
They can also be characterized by their responses to external fields described by the topological field theory \cite{Qi2008}, which is employed in this paper.

In three-dimensional (3D) time-reversal invariant topological insulators, their responses to external fields are described by the so-called theta term \cite{Qi2008}
\begin{align}
\begin{split}
S_\theta&=\frac{e^2}{4\pi^2\hbar c}\int dtd^3 r \theta\bm{E}\cdot\bm{B},
\label{S_theta}
\end{split}
\end{align}
where $\theta=\pi$ (mod $2\pi$), $\bm{E}$ and $\bm{B}$ are an electric field and a magnetic field, respectively.
From this action, we obtain the (quantized) magnetoelectric responses expressed by $\bm{P}= \theta e^2/(4\pi^2\hbar c)\bm{B}$ and $\bm{M}= \theta e^2/(4\pi^2\hbar c)\bm{E}$, with $\bm{P}$ the electric polarization and $\bm{M}$ the magnetization.
While $\theta=\pi$ in the presence of time-reversal symmetry, it is known that the value of $\theta$ can be arbitrary in systems with broken time-reversal symmetry \cite{Essin2009,Coh2011}.
Furthermore, it can be said that the dynamical axion field is realized when $\theta$ is expressed as a function of space and time, $\theta(\bm{r},t)$ \cite{Li2010}.
Note that the axion field can be realized in gapless topological phases such as Weyl semimetals \cite{Zyuzin2012,Son2012}.
Some interesting phenomena arising from the existence of the dynamical axion field have been predicted \cite{Li2010,Zyuzin2012,Son2012,Ooguri2012,Vazifeh2013,Goswami2013,Burkov2015,Chang2015,Buividovich2015,Sekine2015}.

On the other hand, in topological superconductors, their topological nature is captured only by thermal responses \cite{Read2000,Wang2011,Ryu2012}, since charge and spin are not conserved.
It has been heuristically suggested that, in 3D time-reversal invariant (class DIII) topological superconductors and superfluids with topological invariant $N$ \cite{Schnyder2008,Ryu2010}, their responses to external fields are described by the following internal energy term analogous to the theta term \cite{Nomura2012,Shiozaki2013,Comment1}
\begin{align}
\begin{split}
U^g_\theta=\frac{k_B^2 T_0^2}{24\hbar v}\int d^3 r \theta_g\bm{E}_g\cdot\bm{B}_g,
\label{U_theta_g}
\end{split}
\end{align}
where $\theta_g=N\pi$, $T_0$ is the uniform (unperturbed) temperature of the system, and $v$ is the Fermi velocity of surface Majorana fermions.
$\bm{E}_g$ and $\bm{B}_g$ are a {\it gravitoelectric} field and a {\it gravitomagnetic} field, respectively.
As will be explained below, $\bm{E}_g$ ($\bm{B}_g$) corresponds to a temperature gradient (angular velocity vector).
Note that the Lorentz invariance is imposed in the derivation of this term by Nomura {\it et al.} \cite{Nomura2012}.
From this term, we obtain the cross-correlated responses expressed by $\bm{P}_g=\theta_g k_B^2 T_0^2/(24\hbar v)\bm{B}_g$ and $\bm{M}_g=\theta_g k_B^2 T_0^2/(24\hbar v)\bm{E}_g$, with $\bm{P}_g$ the energy polarization and $\bm{M}_g$ the energy magnetization.
As in the case of insulators, the value of $\theta_g$ can be arbitrary in systems with broken time-reversal symmetry, and the realization of the dynamical axion field $\theta_g(\bm{r},t)$ has been suggested \cite{Goswami2014,Shiozaki2014}.
Then what happens when the dynamical axion field is realized in superconductors?
This work is motivated by this question.

In this paper, we search for dynamical cross-correlated responses of 3D topological superconductors and superfluids arising from the existence of the dynamical axion field.
First we derive a general expression for a bulk thermal current in the presence of a temperature gradient and a mechanical rotation (angular velocity).
We show that, in contrast to conventional thermal transport where thermal currents are induced by temperature gradients, a thermal current can be generated by mechanical rotations.
We also show that a thermal current can be generated perpendicular to thermal gradients.
We make a comparison of the cases of superconductors and insulators.
Then we consider a $p$-wave topological superconductor with imaginary $s$-wave pairing fluctuation as a concrete model, focusing on the thermal current generation by mechanical rotations.
Finally, we discuss a possible application of our study to the case of 3D point-nodal superconductors such as Weyl superconductors.

%%%%%%%%%%%
\section{Dynamical Cross-Correlated Responses}
Let us start with introducing a {\it gravitoelectromagnetic} potential $A_g^\mu=(\phi_g,A_g^1,A_g^2,A_g^3)$.
In terms of a gravitoelectromagnetic potential, the gravitoelectric and gravitomagnetic fields are given explicitly by \cite{Mashhoon-Book}
\begin{align}
\begin{split}
\bm{E}_g=-\nabla\phi_g-(1/v)\partial \bm{A}_g/\partial t,\ \ \ \ \ \ \ \bm{B}_g=\nabla\times\bm{A}_g,
\end{split}
\end{align}
where $v$ is the Fermi velocity of a system.
The gravitational potential $\phi_g$ satisfies $\nabla\phi_g=\nabla T/T$ \cite{Luttinger1964}.
Hence, the gravitoelectric field $\bm{E}_g$ can be understood as a temperature gradient.
On the other hand, for example, in a system rotating with the angular velocity $\Omega$ around the $z$ axis, $\bm{A}_g=(A_g^1,A_g^2,A_g^3)$ can be expressed as $\bm{A}_g=(1/v)\Omega\bm{e}_z\times\bm{r}$ \cite{Lynden-Bell1998,Volovik2003}, which gives $\bm{B}_g=(2/v)\Omega\bm{e}_z$.
Therefore, the gravitomagnetic field $\bm{B}_g$ can be understood as an angular velocity vector.
In Ref. \onlinecite{Nomura2012}, a gravitomagnetic field $\bm{B}_g$ was introduced as a quantity which is conjugate to energy magnetization (momentum of energy current) $\bm{M}_g$ in the free energy of a system.
In Lorentz-invariant systems, we have $\bm{M}_g=(v/2)\bm{L}$ with $\bm{L}$ being angular momentum.
Then $\bm{B}_g$ can be understood as the angular velocity vector of a rotating system, which leads to $\bm{B}_g=(2/v)\bm{\Omega}$ \cite{Nomura2012,Volovik2003}.
Here note that, on the other hand, the gravitomagnetic field can also be expressed in terms of a torsion (at least in systems without Lorentz invariance) \cite{Shitade2014,Gromov2015,Bradlyn2015}.
In this formalism, the gravitomagnetic field and the mechanical rotation do not coincide.
Hence, the notation $\bm{B}_g=(2/v)\bm{\Omega}$ should be further examined in the future studies.

In general, $\bm{E}_g$ and $\bm{B}_g$ can depend on time \cite{Mashhoon-Book,Tatara2015}.
Furthermore, as will be discussed below, the fluctuation of $\theta_g$ in a $p+is$-wave superconductor can be written as a function of the relative phase between the two superconducting gaps.
The fluctuation of a relative phase is known as the Leggett mode and can depend on time.
Therefore, it would be appropriate to consider the action of the form \cite{Shiozaki2014}
\begin{align}
\begin{split}
S^g_\theta=\frac{k_B^2 T_0^2}{24\hbar v}\int dt d^3 r \theta_g(\bm{r},t)\bm{E}_g\cdot\bm{B}_g
\label{S_theta_g}
\end{split}
\end{align}
for non-quantized (and dynamical) values of $\theta_g$, instead of the internal energy $U_g^\theta$.
In this paper we call $\theta_g(\bm{r},t)$ the dynamical gravitational axion field.

Now we focus on Lorentz-invariant condensed matter systems where the low-energy effective model is described by relativistic fermions, i.e., Dirac or Majorana fermions.
We introduce a coordinate $x^\mu=(vt,x,y,z)$ with $v$ being the Fermi velocity.
Here note that, as long as we focus on Lorentz-invariant condensed matter systems, the Fermi velocities of the bulk and surface states are the same in the magnitude \cite{CommentA,CommentB}.
In the presence of a weak gravitoelectromagnetic potential $A_g^\mu$, the metric tensor $g_{\mu\nu}$ with $ds^2=g_{\mu\nu}dx^\mu dx^\nu$ is written up to linear order in $A_g^\mu$ as \cite{Mashhoon-Book,Landau-Book}
\begin{align}
\begin{split}
g_{\mu\nu}=
\begin{bmatrix}
1+2\phi_g & A_g^1 & A_g^2 & A_g^3\\
A_g^1 & -1+2\phi_g & 0 & 0\\
A_g^2 & 0 & -1+2\phi_g & 0\\
A_g^3 & 0 & 0 & -1+2\phi_g
\end{bmatrix}.
\end{split}\label{metric-tensor}
\end{align}
The metric is divided into the flat and curved spacetime:
\begin{align}
\begin{split}
g_{\mu\nu}=\eta_{\mu\nu}+h_{\mu\nu}
\end{split}
\end{align}
with $\eta_{\mu\nu}=\mathrm{diag}(+1,-1,-1,-1)$ and $h_{\mu\nu}\ll 1$.
All the subscripts and superscripts in this paper are raised and lowered by $\eta^{\mu\nu}$ and $\eta_{\mu\nu}$, respectively.
Note that $g^{\mu\nu}=\eta^{\mu\nu}-h^{\mu\nu}$ ($h^{\mu\nu}=\eta^{\mu\alpha}\eta^{\nu\beta}h_{\alpha\beta}$) due to the identity $g_{\mu\nu}g^{\nu\rho}={\delta_\mu}^\rho$.

The energy-momentum tensor $T^{\mu\nu}$ of a Lorentz-invariant system is obtained from the variation of the action with respect to the metric tensor.
In systems whose effective action is given by Eq. (\ref{S_theta_g}), we have
\begin{align}
\begin{split}
T^{\mu\nu}=\frac{2}{\sqrt{-g}}\frac{\delta S_\theta^g}{\delta g_{\mu\nu}},
\end{split}
\end{align}
where $\sqrt{-g}=\sqrt{-\mathrm{det}(g)}\simeq 1+h/2$ with $h={h^\mu}_\mu$.
The energy current density $j_E^a$ ($a=1,2,3$) is given by $j_E^a=vT^{0a}$.
After a straightforward calculation from Eq. (\ref{S_theta_g}), we obtain
\begin{align}
\begin{split}
\frac{\delta S_\theta^g}{\delta A_g^a}=\frac{k_B^2 T_0^2}{24\hbar v}\left[\frac{1}{v}\dot{\theta}_g(\bm{r},t) B_g^a+(\nabla\theta_g(\bm{r},t)\times\bm{E}_g)^a\right],
\end{split}
\end{align}
where $\dot{\theta}_g=\partial \theta_g(\bm{r},t)/\partial t$.
The thermal current $\bm{j}_T$ is given by $\bm{j}_T=\bm{j}_E-(\mu/e)\bm{j}$ with $\bm{j}$ an electric current and $\mu$ the quasiparticle chemical potential.
In superconductors described by the Bogoliubov-de Gennes Hamiltonian, $\mu=0$ is always satisfied due to particle-hole symmetry, which lead to $\bm{j}_T=\bm{j}_E$.
Finally, we obtain a general expression for the thermal current in the bulk
\begin{align}
\begin{split}
\bm{j}_T(\bm{r},t)=\frac{k_B^2 T_0^2}{12\hbar v}\left[\dot{\theta}_g(\bm{r},t) \bm{B}_g+v\nabla\theta_g(\bm{r},t)\times\bm{E}_g\right],\label{thermalcurrent}
\end{split}
\end{align}
where we have neglected the $\mathcal{O}(h^2)$ terms, since $\bm{B}_g$ and $\bm{E}_g$ are both $\mathcal{O}(h_{\mu\nu})$ and we have considered the case of $h_{\mu\nu}\ll 1$.
Note that this thermal current does not flow when $\theta_g$ takes constant values, i.e., when $\dot{\theta}_g=\nabla\theta_g=0$.
In other words, the generation of the thermal current (\ref{thermalcurrent}) is a consequence of the realization of the dynamical gravitational axion field $\theta_g(\bm{r},t)$.
The first term in the bracket indicates that a thermal current is induced in the bulk of a 3D superconductor by gravitomagnetic fields, i.e., by mechanical rotations.
We call this effect the {\it chiral gravitomagnetic effect} \cite{Landsteiner2011}.
The second term in the bracket indicates that a thermal current is induced in the bulk by gravitoelectric fields, i.e., by temperature gradients.
This current is perpendicular to temperature gradients, and thus this is the anomalous thermal Hall effect, where ``anomalous'' means the absence of magnetic fields.

We note that the chiral gravitomagnetic effect and the anomalous thermal Hall effect we have derived so far can be understood as a ``polarization current'' in the bulk and a ``magnetization current'' in the bulk \cite{Qin2011}, respectively, which are obtained from the internal energy (\ref{U_theta_g}).
Namely, they can be obtained from $\partial\bm{P}_g/\partial t=k_B^2 T_0^2/(24\hbar v)\dot{\theta}_g\bm{B}_g$ and $v\nabla\times\bm{M}_g=k_B^2 T_0^2/(24\hbar)\nabla\theta_g\times\bm{E}_g$, apart from the factor $2$.

\begingroup
\renewcommand{\arraystretch}{1.6}
\begin{table*}[!t]
\caption{Comparison of dynamical cross-correlated responses of insulators and superconductors.
In 3D insulators (or semimetals) whose effective action is given by $S_\theta$, the chiral magnetic effect, an electric current generation by magnetic fields, and the anomalous Hall effect arise in the bulk due to the presence of the dynamical axion field $\theta(\bm{r},t)$ [see Eq. (\ref{insulators-current})].
In 3D superconductors (or superfluids) whose effective action is given by $S^g_\theta$, the chiral gravitomagnetic effect, a thermal current generation by gravitomagnetic fields, i.e, by mechanical rotations, and the anomalous thermal Hall effect arise in the bulk due to the presence of the dynamical gravitational axion field $\theta_g(\bm{r},t)$ [see Eq. (\ref{thermalcurrent})].}
\begin{ruledtabular}
\begin{tabular}{ccc}
Effective action & Magnetic field $\bm{B}$ (Gravitomagnetic field $\bm{B}_g$) & Electric field $\bm{E}$ (Gravitoelectric field $\bm{E}_g$)\\
\hline
$S_\theta=\frac{e^2}{4\pi^2\hbar c}\int dtd^3 r \theta\bm{E}\cdot\bm{B}$ & 
Chiral magnetic effect\ \ \ $\bm{j}\propto \dot{\theta}\bm{B}$ & Anomalous Hall effect\ \ \ $\bm{j}\propto \nabla\theta\times\bm{E}$\\
$S^g_\theta=\frac{k_B^2 T_0^2}{24\hbar v}\int dtd^3 r \theta_g\bm{E}_g\cdot\bm{B}_g$ & Chiral gravitomagnetic effect\ \ \ $\bm{j}_T\propto \dot{\theta}_g\bm{B}_g$ & Anomalous thermal Hall effect\ \ \ $\bm{j}_T\propto \nabla\theta_g\times\bm{E}_g$
\end{tabular}
\end{ruledtabular}\label{Table1}
\end{table*}
\endgroup
Here let us consider a similarity between the cases of superconductors and insulators (or semimetals).
In 3D insulators (or semimetals), it is known that an electric current is derived from the theta term (\ref{S_theta}) as $j^a=\delta S_\theta/\delta A_a$ ($A_a$ is the spatial component of an electromagnetic potential) \cite{Zyuzin2012,Son2012,Goswami2013,Sekine2015}:
\begin{align}
\begin{split}
\bm{j}(\bm{r},t)=\frac{e^2}{4\pi^2 \hbar c}\left[\dot{\theta}(\bm{r},t)\bm{B}+\nabla\theta(\bm{r},t)\times\bm{E}\right].\label{insulators-current}
\end{split}
\end{align}
The magnetic-field-induced term is the so-called chiral magnetic effect \cite{Fukushima2008}, and the electric-field-induced term is the anomalous Hall effect.
These two effects occur as a consequence of the chiral anomaly and the realization of the dynamical axion field $\theta(\bm{r},t)$.
The electric current of the form (\ref{insulators-current}) has been studied in Weyl semimetals \cite{Zyuzin2012,Vazifeh2013,Goswami2013,Burkov2015,Chang2015,Buividovich2015} and antiferromagnetic insulators with spin-orbit coupling \cite{Sekine2015}.
From Eqs. (\ref{thermalcurrent}) and (\ref{insulators-current}), we see that there is an analogy such that
\begin{align}
\begin{split}
\dot{\theta}_g(\bm{r},t) \bm{B}_g\ &\leftrightarrow\ \dot{\theta}(\bm{r},t)\bm{B},\\
\nabla\theta_g(\bm{r},t)\times\bm{E}_g\ &\leftrightarrow\ \nabla\theta(\bm{r},t)\times\bm{E}.
\end{split}
\end{align}
Table \ref{Table1} shows a comparison of dynamical cross-correlated responses of insulators and those of superconductors.

%%%%%%%%%%%
\subsection{Chiral Gravitomagnetic Effect}
To see the emergence of the chiral gravitomagnetic effect [the first term in Eq. (\ref{thermalcurrent})], let us consider a concrete model for a 3D topological superconductor (or superfluid).
In time-reversal invariant 3D topological superconductors with topological invariant $N$, $\theta_g$ in Eq. (\ref{U_theta_g}) takes quantized value $N\pi$.
To induce the deviation of $\theta_g$ from the quantized value, time-reversal symmetry of the bulk needs to be broken, as in the case of insulators.
It has been shown that the imaginary $s$-wave pairing fluctuation in class DIII topological superconductors such as the $^3$He-B phase leads to the deviation of the value of $\theta_g$ from $\pi$ \cite{Goswami2014,Shiozaki2014}.
Here such an imaginary $s$-wave pairing order corresponds to the chiral symmetry breaking term (which also breaks time-reversal symmetry) $\Gamma=\Theta\Xi$, with $\Theta$ and $\Xi$ being the time-reversal and particle-hole operators, respectively \cite{Shiozaki2014,Wang2011,Shiozaki2013}.
Thus the resulting superconducting state belongs to the class D \cite{Schnyder2008,Ryu2010,Altland1997}.
Furthermore, in class DIII ($p$-wave) topological superconductors with imaginary $s$-wave fluctuation, the dynamical gravitational axion field $\delta\theta_g(\bm{r},t)$ can be realized by the relative phase fluctuation, i.e., the Leggett mode \cite{Goswami2014,Shiozaki2014}.

The Bogoliubov-de Gennes Hamiltonian we consider is given by $H=\sum_{\bm{k}}\Psi^\dag_{\bm{k}}\mathcal{H}_{\rm BdG}(\bm{k})\Psi_{\bm{k}}$, where
\begin{align}
\begin{split}
\mathcal{H}_{\rm BdG}(\bm{k})&=
\begin{bmatrix}
\xi_{\bm{k}} & i\Delta_s^{\rm Im}+(\Delta_p/k_F)\bm{k}\cdot\bm{\sigma}\\
-i\Delta_s^{\rm Im}+(\Delta_p/k_F)\bm{k}\cdot\bm{\sigma} & -\xi_{\bm{k}}
\end{bmatrix}\label{BdG}
\end{split}
\end{align}
and $\Psi^\dag_{\bm{k}}=(c^\dag_{\bm{k}\uparrow}, c^\dag_{\bm{k}\downarrow}, c_{-\bm{k}\downarrow}, -c_{-\bm{k}\uparrow})$ is a Nambu spinor.
Here $k_F$ is the Fermi wave number, $\xi_{\bm{k}}=\frac{k^2}{2m}-\mu$ with $\mu$ the chemical potential, $\Delta_s^{\rm Im}$ and $\Delta_p$ are the imaginary $s$-wave and $p$-wave pairing amplitudes, respectively, and $\bm{\sigma}=(\sigma_1,\sigma_2,\sigma_3)$ are the Pauli matrices for the spin space.
Defining $4\times4$ matrices $\alpha_\mu$ that satisfy the Clifford algebra $\{\alpha_\mu,\alpha_\nu\}=2\delta_{\mu\nu}$ with $\alpha_5=\alpha_1\alpha_2\alpha_3\alpha_4$, Eq. (\ref{BdG}) can be rewritten as
\begin{align}
\begin{split}
\mathcal{H}_{\rm BdG}(\bm{k})=(\Delta_p/k_F)\bm{k}\cdot\bm{\alpha}+\xi_{\bm{k}}\alpha_4+\Delta_s^{\rm Im}\alpha_5.
\end{split}
\end{align}
Clearly this is a massive Dirac Hamiltonian with the mass term $\Delta_s^{\rm Im}\alpha_5$ which breaks time-reversal and inversion symmetries.
When $\mu>0$ ($\mu<0$) with $\Delta_s^{\rm Im}=0$, the system is topologically nontrivial (trivial) \cite{Schnyder2008,Qi2009}.
As mentioned above, the imaginary $s$-wave pairing results in the non-quantized value of $\theta_g$ such that $\theta_g=\pi+\mathrm{tan}^{-1}(\Delta^{\rm Im}_s/\mu)$ \cite{Goswami2014}.
This is because the imaginary $s$-wave pairing is exactly the same as the $\gamma_5$ (or $\alpha_5$ here) matrix.
As in the case of insulators \cite{Li2010,Sekine2014}, the $\gamma_5$ matrix generates the deviation of $\theta_g$ from $\pi$.
When we take into account the superconducting fluctuations $\Delta^{\rm Im}_s=|\Delta^{\rm Im}_s|e^{i\theta_s(\bm{r},t)}$ and $\Delta_p=|\Delta_p|e^{i\theta_p(\bm{r},t)}$, the relative phase fluctuation $\theta_r(\bm{r},t)\equiv \theta_s(\bm{r},t)-\theta_t(\bm{r},t)$, i.e., the Leggett mode gives rise to the dynamical gravitational axion field, as $\delta\theta_g(\bm{r},t)\propto\delta\theta_r(\bm{r},t)$ \cite{Goswami2014,Shiozaki2014}.

\begin{figure}[!t]
\centering
\includegraphics[width=0.7\columnwidth,clip]{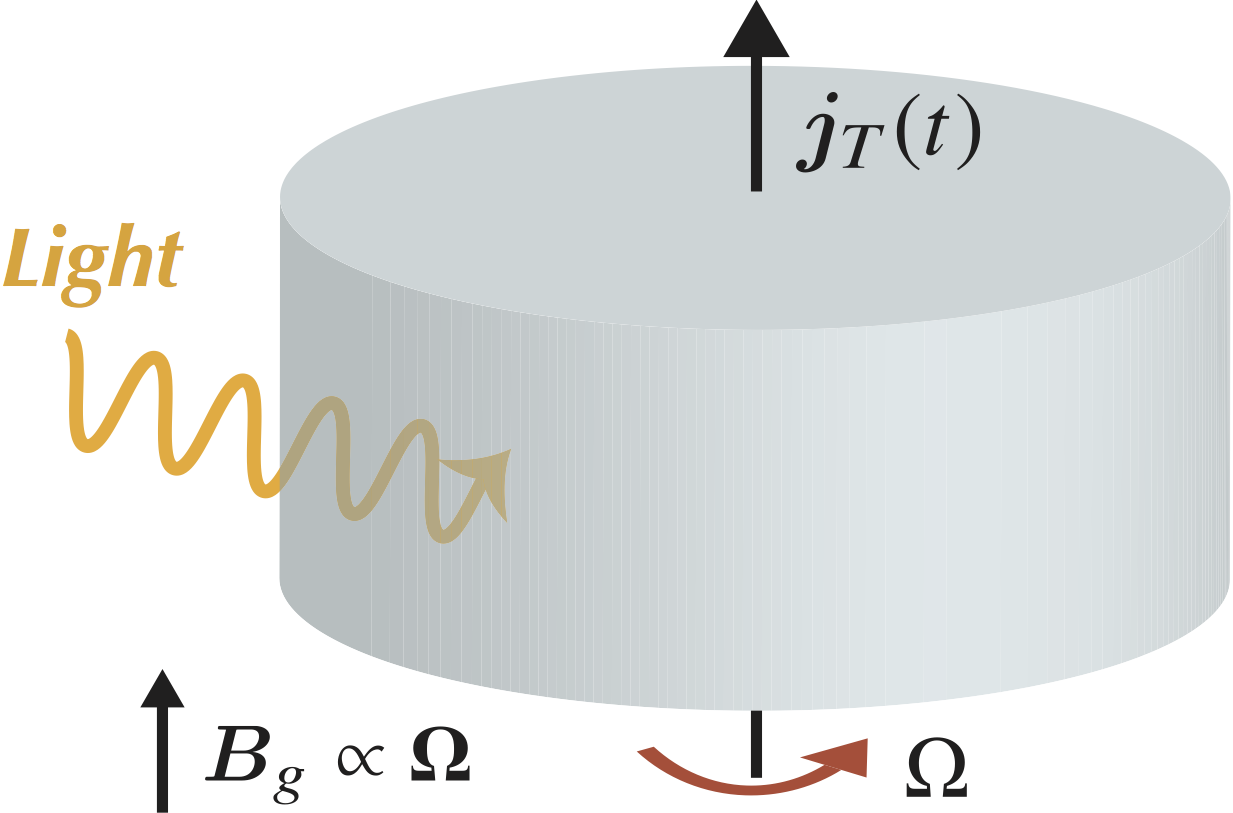}
\caption{(Color online) Schematic figure of the chiral gravitomagnetic effect in 3D class DIII topological superconductors and superfluids with imaginary $s$-wave pairing fluctuation.
Irradiation of light excites the Leggett mode, which gives rise to the dynamical gravitational axion field $\theta_g(t)$.
In the presence of the dynamical gravitational axion field, an alternating thermal current $\bm{j}_T(t)$ is induced by a static gravitomagnetic field $\bm{B}_g$, i.e., by a mechanical rotation with constant frequency $\Omega$.}\label{Fig1}
\end{figure}
The Leggett modes can be excited by electric fields (i.e., light) through the coupling of photons and the charge density \cite{Devereaux2007,Blumberg2007}.
Here let us consider a case where the resonance state, i.e., the uniform ($\bm{q}=0$) Leggett mode is realized.
In this case, the relative phase fluctuation can be written as
\begin{align}
\begin{split}
\theta_r(t)=\theta_r^0+\delta\theta_{L}\sin(\omega_Lt+\alpha),\label{Leggett-resonance}
\end{split}
\end{align}
where $\theta_r^0$ is the mean-field value, $\delta\theta_{L}(\sim 0.01\pi)$ is the amplitude of the fluctuation \cite{Comment2}, $\omega_L$ is the resonance frequency, and $\alpha$ is a constant.
Combining the first term in Eq. (\ref{thermalcurrent}) and Eq. (\ref{Leggett-resonance}), we obtain an explicit expression for the chiral gravitomagnetic effect
\begin{align}
\begin{split}
\bm{j}_T(t)=\frac{k_B^2 T_0^2}{6\hbar v^2}D\delta\theta_{L}\omega_L\bm{\Omega}\cos(\omega_Lt+\alpha),
\end{split}
\end{align}
where we have used $\bm{B}_g=(2/v)\bm{\Omega}$ with $\bm{\Omega}$ being an angular velocity vector, and $\delta\theta_g=D\delta\theta_r$ with $D\sim 1$ \cite{Shiozaki2014}.
This equation indicates that an alternating thermal current is induced by mechanical rotations, when the Leggett mode is excited. 
Figure \ref{Fig1} shows a schematic diagram to realize the chiral gravitomagnetic effect in our system.

Here let us estimate the maximum value of the chiral gravitomagnetic effect: $|\bm{j}_T|_{\rm max}=\frac{k_B^2 T_0^2}{6\hbar v^2}D\delta\theta_{L}\omega_L\Omega$.
Theoretically, Cu$_x$Bi$_2$Se$_3$ can be a class DIII topological superconductor \cite{Fu2010}.
On the other hand, recent experiments on Cu$_x$Bi$_2$Se$_3$ have suggested two scenarios: it is a topological superconductor \cite{Wray2010,Sasaki2011,Kirzhner2012} or a conventional $s$-wave superconductor \cite{Levy2013,Peng2013}.
These experimental results may indicate a possible competition between triplet and singlet pairings.
The gravitational axion mass gap $\hbar\omega_L$ should be smaller than the superconducting gap.
From an experimental value of the superconducting gap in Cu$_x$Bi$_2$Se$_3$ ($\Delta\sim 0.7\ \mathrm{meV}$ at $T_0=0\ \mathrm{K}$) \cite{Kriener2011}, we assume $\hbar\omega_L\sim 0.1\ \mathrm{meV}$ (or $\omega_L\sim 100\ \mathrm{GHz}$).
Also, we use experimental values of Cu$_x$Bi$_2$Se$_3$ such that $T_0\sim 3\ \mathrm{K}$ and $v\sim 5\times 10^5\ \mathrm{m/s}$ \cite{Wray2010,Hor2010}.
Furthermore, from the condition $A_g^i\ll 1$, we assume $\Omega\sim 10^4\ \mathrm{Hz}$.
Substituting these possible values, we obtain $|\bm{j}_T|_{\rm max}\sim 1\times 10^{-9}\ \mathrm{W/m^2}$.
It would be possible to observe this value experimentally.

%%%%%%%%%%%
\subsection{Anomalous Thermal Hall Effect}
Next we consider briefly the realization of the anomalous thermal Hall effect [the second term in Eq. (\ref{thermalcurrent})] in the bulk of a $p+is$ superconductor described by Eq. (\ref{BdG}).
Let us suppose a case where a gravitoelectric field $\bm{E}_g=-\bm{e}_y\partial_yT/T$ is applied, and there is a spatial variation (domain wall) of the relative phase $\theta_r(\bm{r})$ in the $z$ direction such that $\theta_r(z=0)=\theta^0_r$ and $\theta_r(z=L)=\theta^0_r+\eta$.
Then we obtain $j^x_T(z)=\frac{k_B^2 T_0}{12\hbar}D\left(\partial_z\theta_r(z)\right)\partial_yT$.
The total thermal current in the $x$ direction is given by
\begin{align}
\begin{split}
J^x_T=\int_0^L dzj^x_T(z)=\frac{k_B^2 T_0}{12\hbar}D\eta\partial_yT.
\end{split}
\end{align}
The thermal Hall conductivity $\lambda_{xy}=\frac{k_B^2 T_0}{12\hbar}D\eta$ can be written in terms of the quantized thermal Hall conductivity of surface Majorana fermions $\kappa_{xy}=\frac{\pi k_B^2 T_0}{24\hbar}$ \cite{Read2000,Wang2011,Ryu2012,Nomura2012,Shiozaki2013} as $\lambda_{xy}=\frac{2D\eta}{\pi}\kappa_{xy}$.
Since $D\sim 1$ and $\eta\sim\pi$, we get $\lambda_{xy}\sim\kappa_{xy}$.
Experimental observation of this thermal Hall effect seems, however, to be difficult.
The problem is that the realization of the spatial variation of the relative phase $\theta_r(\bm{r})$ is not easy.
In the case of single superconducting gap systems, we can use a Josephson junction.
However, spatial control of the relative phase between two superconducting gaps would be difficult.

%%%%%%%%%%%
\section{Discussions}
So far we have considered the dynamical realization of the chiral gravitomagnetic effect.
Is it possible to realize the chiral gravitomagnetic effect without collective excitations by external forces such as the Leggett mode?
Based on the analogy with Weyl semimetals, let us consider this possibility.
In the case of Weyl semimetals, $\theta(\bm{r},t)$ is given by $\theta(\bm{r},t)=2(\bm{b}\cdot\bm{r}-b_0t)$, where $\bm{b}$ is the distance between the two Weyl nodes in momentum space and $b_0$ is the chemical potential difference between the nodes \cite{Zyuzin2012,Vazifeh2013,Goswami2013,Burkov2015}.
Then we obtain the electric current as $\bm{j}_{\rm WSM}=-\frac{e^2}{2\pi^2 \hbar c}b_0\bm{B}+\frac{e^2}{2\pi^2 \hbar c}\bm{b}\times\bm{E}$, which means that the current can be induced in the ground states.
Hence the induced current $\bm{j}_{\rm WSM}$ can be regarded as a {\it static} response.
A naive presumption may allow us to expect that in a 3D point-nodal superconductor $\theta_g(\bm{r},t)$ takes the form
\begin{align}
\begin{split}
\theta_g(\bm{r},t)=2(\bm{b}_g\cdot\bm{r}-b_{g0}t),\label{theta_g-WeylSC}
\end{split}
\end{align}
where $\bm{b}_g$ is the distance between the two point nodes in the superconducting gap in momentum space, and $b_{g0}$ is the chemical potential difference between the nodes.
If this is true, then we obtain the induced thermal current as
\begin{align}
\begin{split}
\bm{j}_T=-\frac{k_B^2 T_0^2}{6\hbar v}b_{g0}\bm{B}_g+\frac{k_B^2 T_0^2}{6\hbar}\bm{b}_g\times\bm{E}_g.\label{thermal-current-WeylSC}
\end{split}
\end{align}
Note that, although we could consider at least theoretically the case of nonzero $b_{g0}$, it will be impossible that values of $b_{g0}$ become nonzero in realistic systems, as in the case of Weyl semimetals.
Here let us consider a possible application to the case of 3D Weyl superconductors \cite{Yang2014}.
A recent study have suggested that, in 3D superconductors with two Weyl nodes in the $k_z$ axis, the anomalous thermal Hall conductivity is given by \cite{Goswami2015}
\begin{align}
\begin{split}
\kappa_{xy}=\frac{\pi k_B^2 T_0}{6\hbar}C(k_z)\frac{\Delta k_z}{2\pi},\label{kappa_xy-WeylSC}
\end{split}
\end{align}
where $C(k_z)$ is the Chern number with $k_z$ fixed between the Weyl nodes, and $\Delta k_z$ is the distance between the Weyl nodes in momentum space.
We can reproduce Eq. (\ref{kappa_xy-WeylSC}) by setting $\bm{b}_g=\frac{1}{2}C(k_z)\Delta k_z\bm{e}_z$ and $\bm{E}_g=-\bm{e}_y\partial_yT/T$ in Eq. (\ref{thermal-current-WeylSC}).
However, further studies are needed to verify the applicability of our prediction (\ref{theta_g-WeylSC}) to Weyl superconductors.

We would like to comment on the relation of the chiral gravitomagnetic effect in our study with the ``chiral vortical effect'' \cite{Son2009,Erdmenger2009,Landsteiner2011a}.
The chiral vortical effect is commonly referred to as an electric current generation by vorticity (i.e., rotation) in gapless fermion systems.
In a broad sense, the chiral vortical effect includes an energy current generation by vorticity \cite{Landsteiner2011,Landsteiner2013}.
Such an energy current generation has been discussed in Weyl semimetals \cite{Landsteiner2014}.
The chiral vortical effect has contributions from the chiral anomaly and the gravitational anomaly.
It should be noted that the contribution from the gravitational anomaly have a $T^2$ dependence ($T$ is a temperature) \cite{Landsteiner2011a,Landsteiner2013}.
On the other hand, the thermal current (\ref{thermalcurrent}) comes only from the gravitational topological term (\ref{U_theta_g}).
In addition, the relation between the gravitational topological term (\ref{U_theta_g}) and the gravitational anomaly is unclear at present.
Therefore, based on the comparison of Eqs. (\ref{thermalcurrent}) and (\ref{insulators-current}), we call a thermal current generation by mechanical rotations the chiral gravitomagnetic effect.

%%%%%%%%%%%
\section{Summary}
In summary, we have derived dynamical cross-correlated responses of 3D topological superconductors and superfluids from a gravitational topological term analogous to the theta term.
We have proposed the emergence of the ``chiral gravitomagnetic effect'', a thermal current generation by gravitomagnetic fields, i.e., by mechanical rotations.
This effect is in contrast to conventional thermal transport where thermal currents are induced by temperature gradients, and can be regarded as a thermal counterpart of the chiral magnetic effect.
We have shown that, in class DIII topological superconductors with imaginary $s$-wave pairing fluctuation, the chiral gravitomagnetic effect can be realized dynamically due to the presence of the time dependences of the relative phase, i.e., the Leggett mode.
We have also shown the occurrence of the anomalous thermal Hall effect in the bulk due to the presence of the spatial variations of the relative phase.
We have briefly discussed a possible application of our study to the thermal responses of 3D point-nodal superconductors such as Weyl superconductors.
These two phenomena arise as a consequence of the realization of the dynamical gravitational axion field in condensed matter.
Finally, we should note that Eq. (\ref{U_theta_g}), which is our starting point, has not yet been derived microscopically as a bulk quantity, i.e., without referring to the existence of surface states.
Microscopic derivation of Eq. (\ref{U_theta_g}) is an important open issue.

%%%%%%%%%%%
\acknowledgements
The author thanks K. Nomura, R. Nakai, Y. Araki, and T. Koyama for valuable discussions.
The author is supported by a JSPS Research Fellowship.

%\appendix


\begin{thebibliography}{99}
\bibitem{CommentB}
More generally, this can be justified by the following argument.
Topological properties of a system are not changed qualitatively by continuous parameter changes, i.e., they do not depend on the detail of the electronic band structure, as long as the bulk topological invariant is retained, which means that the effective Hamiltonian of the bulk can be deformed so that the Fermi velocities of the bulk and surface states are the same in the magnitude.
\bibitem{Comment2}
Although it is difficult to estimate the amplitude of the relative phase fluctuation $\delta\theta_L$, it will be much smaller than the mean-field (ground-state) value $\theta^0_r$ that is of the order of $\pi$.
We assume that $\delta\theta_L\sim 0.01\pi$ in this study.
\end{thebibliography}
\end{document}